\documentclass{article}
\usepackage{amssymb}

\usepackage{graphicx}

\input{tcilatex}

\begin{document}

\title{Plasma Frequency Shift Due to a Slowly Rotating Compact Star}
\author{Babur M. Mirza \\
{\small Department of Mathematics, Quaid-i-Azam University, }\\
{\small Islamabad 45320, Pakistan}\\
Hamid Saleem \and {\small Theoretical Physics Group, PINSTECH, Nilore,
Islamabad, Pakistan; and } \and {\small COMSATS Institute of Information
Technology, F-8, Islamabad, Pakistan}}
\date{May 10, 2005}
\maketitle

\begin{abstract}
We investigate the effects of a slowly rotating compact gravitational source
on electron oscillations in a homogeneous electrically neutral plasma in the
absence of an external electric or magnetic field. Neglecting the random
thermal motion of the electrons we assume the gravitoelectromagnetic
approximation to the general theory of relativity for the gravitational
field. It is shown that there is a shift in the plasma frequency and hence
in the dielectric constant of the plasma due to the gravitomagnetic force.
We also give estimates for the difference in the frequency of radially
transmitted electromagnetic signals for typical compact star candidates.
\end{abstract}

\section{Introduction}

Compact stars consist mostly of a degenerated plasma which close to the
surface of the star forms the so called ion crust of densely packed ions and
relatively free electrons. Moreover a highly ionized plasma constitutes the
stellar atmosphere formed as a result of ionization due to the increase in
the mean collisional rate of the atoms constituting the star's atmosphere
\cite{[1]}. Outside the ion crust the plasma co-rotating with the star is
rather dilute and highly conducting\cite{[2]}. A plasma, both in a
degenerate form and as a dilute and highly conducting medium, behaves as an
oscillating system having a characteristic frequency called the plasma
frequency. In particular the dielectric constant of a medium is determined
by the plasma frequency, which in turn determines the transmission and
reflection of electromagnetic radiation for the plasma\cite{[3]}.

On the other hand compact stars possess very strong gravitational fields, so
that general relativistic effects are important for an adequate description
of the phenomenon occuring in vicinity of these stars. Usually general
relativistic effects are not directly observable but are manifest in an
indirect way, for example via interaction with the magnetic field\cite{[4],
[5]}, in the accretion of matter\cite{[6], [7], [8], [9]}, and other
material and radiative processes occurring in vicinity of the star (e.g.
Ref. [2]). An investigation of these effects is of interest for various
astrophysical processes as well as for testing the general theory of
relativity\cite{[10]}.

In this paper we investigate gravitomagnetic effects of a slowly rotating
compact gravitational source on the electron oscillations in a homogeneous
electrically neutral plasma. We assume the absence of any electric or
magnetic field and neglect the random thermal motion of the electrons. In
the next section we present the gravitoelectromagnetic approximation to the
general theory of relativity particularly for the geodesic equation. In
section 3 we formulate the equations of motion for the electron oscillations
in the gravitational field of the star and study the effects on plasma
frequency. It found that there is a reduction in the plasma frequency due to
the gravitomagnetic force. We then estimate in section 4 the effects of the
variation in the dielectric constant on the electromagnetic waves
propagating radially through the plasma surrounding the compact
gravitational source. In conclusion we summarize the main results of the
paper and discuss their relevance to observation. Throughout we use the
gravitational units $G=1=c$ unless mentioned otherwise.

\section{The Gravitoelectromagnetic Approximation}

The field equations of general theory of relativity, for a slowly rotating
gravitational source, bear a remarkable formal similarity with the
fundamental equations of classical electromagnetism\cite{[11], [12]}.
Particularly in Einstein's theory of gravitation the trajectory of a test
particle in vicinity of a gravitational source is given by the geodesic
equation:

\begin{equation}
\frac{d^{2}x^{\alpha }}{ds^{2}}+\Gamma _{\beta \gamma }^{\alpha }\frac{%
dx^{\beta }}{ds}\frac{dx^{\gamma }}{ds}=0;
\end{equation}
For sufficiently weak gravitating systems, such as the compact stars, a
linearization of the metric can be adequately assumed, where

\begin{equation}
g_{\alpha \beta }=\eta _{\alpha \beta }+h_{\alpha \beta },
\end{equation}
and where $\eta _{\alpha \beta }$ $=diag(-1,-1,-1,1)$ is the Minkowski
metric tensor and $h_{\alpha \beta }$ is the perturbation to the metric such
that $h_{\alpha \beta }\ll 1$ and $x^{\alpha }$ $\equiv (\mathbf{x}%
^{i},x^{0})=(\mathbf{r,}t)$ are the position coordinates of test particle,
with time $t$ being the affine parameter, the geodesic equation takes the
form 
\begin{equation}
\frac{d^{2}\mathbf{r}}{dt^{2}}=\mathbf{G}+\mathbf{v}\times \mathbf{H,}
\end{equation}
where 
\begin{equation}
\mathbf{G=-\nabla }\varphi ,\quad \mathbf{H=\nabla \times }4\mathbf{a,}
\end{equation}
and

\begin{equation}
\varphi =-\iiint \frac{\rho }{r}dV,\quad \mathbf{a=}\iiint \frac{\rho 
\mathbf{v}}{r}dV.
\end{equation}
Assuming to the first order of approximation, the star to be a slowly
rotating sphere of homogeneous mass $M$ and radius $R$, we have 
\begin{equation}
\mathbf{G}=-M\widehat{\mathbf{e}}_{\mathbf{r}}/r^{2},
\end{equation}
which is the gravitoelectric force given by the Newtonian gravitational
force per unit mass, and $\widehat{\mathbf{e}}_{\mathbf{r}}$ is a unit
vector in the radial direction. The term $m($ $\mathbf{v\times H)}$\textbf{\ 
}is the gravitomagnetic force where $\mathbf{H}$ is given by

\begin{equation}
\mathbf{H}=-\frac{12}{5}MR^{2}(\mathbf{\Omega .r}\frac{\mathbf{r}}{r^{5}}-%
\frac{1}{3}\frac{\mathbf{\Omega }}{r^{3}}),
\end{equation}
$\mathbf{\Omega }$ being the angular frequency vector of the gravitational
source, and $\mathbf{v}$ is velocity of the test particle. The
approximation, called the gravitoelectromagnetic approximation to the
general theory of relativity, is generally valid for compact astrophysical
sources. The independence of the gravitomagnetic potential $\mathbf{a}$ from
a particular frame and particular coordinate system has been demonstrated 
\cite{[13]}. Physically it can be interpreted as `gravitomagnetic current'
induced in the vicinity of the gravitational source due to its rotation.

\section{Equations of Motion for Electron Oscillations}

With a fixed ion background it is convenient to choose an orientation of the
coordinates triplet $(x,y,z)$ such that the angular frequency vector $%
\mathbf{\Omega }$ is along the positive $z$-axis. Since the electron
oscillations occur within a very small region of space we have, for the
components of acceleration due to gravitoelectric force, $\mathbf{G}$ to be
of constant magnitude $g$ along each direction $(x,y,z)$. In this case the
components of $\mathbf{H}$ are $(0,0,H)$, therefore effective components of
the gravitomagnetic force lie in the $xy$-plane. We take, without loss of
generality, the $xy$-plane to be the equatorial plane of the star mainly
because here the gravitomagnetic effects are of maximum magnitude for the
given orientation\cite{[14], [15]}.

The equations of motion for the electron oscillations are:

\begin{equation}
\frac{d^2x}{dt^2}=-\omega _p^2x-g-H\frac{dy}{dt},
\end{equation}

\begin{equation}
\frac{d^{2}y}{dt^{2}}=-\omega _{p}^{2}y-g+H\frac{dx}{dt},
\end{equation}
where $\omega _{p}=\sqrt{N_{0}e^{2}/m_{e}\epsilon _{0}}$ is the Newtonian
plasma frequency, $N_{0}$ being the electron number density of the plasma
per centimeter, $e$ is the electronic charge, $m_{e}$ is the electronic
mass, and $\epsilon _{0}$ is the dielectric constant. Clearly $H$ has
dimensions of cycle per unit time i.e. of frequency.

To investigate the resonant frequency for the above system of coupled
equations, we assume that the solutions to equations (8), and (9) can be
expressed as $x=a\exp (i\omega t)-g/\omega _{p}^{2}$ and $y=b\exp (i\omega
t)-g/\omega _{p}^{2}$ where $\omega $ is the applied external frequency. The
system can then be written as a single matrix equation

\begin{equation}
\left[ 
\begin{array}{cc}
\omega ^{2}-\omega _{p}^{2} & -i\omega H \\ 
i\omega H & \omega ^{2}-\omega _{p}^{2}
\end{array}
\right] \left[ 
\begin{array}{c}
a \\ 
b
\end{array}
\right] =0.
\end{equation}
For solution to exist equation (10) implies that,

\begin{equation}
\left| 
\begin{array}{cc}
\omega ^2-\omega _p^2 & -i\omega H \\ 
i\omega H & \omega ^2-\omega _p^2
\end{array}
\right| \equiv 0.
\end{equation}

Since $H\ll \omega _p$, we expand the determinant and neglect terms
involving squares and higher powers of $H/\omega _p$. This gives the
following expression for resonant frequency of the oscillating plasma

\begin{equation}
\omega \simeq \omega _{p}-H/2.
\end{equation}
Further if $\chi $ is the angle between the position vector $\mathbf{r}$ and
the angular frequency vector $\mathbf{\Omega }$, it follows from expression
(7) that close to the surface of the star, i. e., at $r\simeq R$ , $\mathbf{H%
}$ has magnitude given by

\begin{equation}
H\equiv \mid \mathbf{H}\mid \simeq \mu \sqrt{1+3\cos ^2\chi },
\end{equation}
where $\mu =(4GM/5Rc^2)\Omega $, $\Omega $ being the magnitude of the
angular frequency vector.

Substituting from expression (13) in (12) we obtain an expression relating
plasma frequency $\omega _{p}$ of the star's atmosphere to the angle of
inclination $\chi $:

\begin{equation}
\frac{\omega }{\omega _{p}}\simeq 1-\frac{\mu }{2\omega _{p}}\sqrt{1+3\cos
^{2}\chi },\quad \mu \ll \omega _{p}.
\end{equation}
In expression (14) we see that the shift in plasma frequency depends not
only on the mass, radius and angular frequency of the compact star, via the
parameter $\mu $, but also on the angle of inclination $\chi $. Here the
parameter $\mu $ determines the magnitude of the gravitomagnetic effect for
different compact stellar sources. For a typical compact star $\mu $ ranges
from $0.1687\times 10^{-3}Hz$ (for a typical white dwarf of mass $%
1M_{\circledast }=1.989\times 10^{30}kg$ , radius $7\times 10^{6}m$ and
angular frequency$1Hz$) to $236.2932Hz$ (for a typical neutron star of mass $%
2M_{\circledast }$, radius $1\times 10^{4}m$, and angular frequency $1kHz$)
with corresponding plasma frequency ranging approximately $5.65\times
10^{2}Hz$ to $5.65\times 10^{6}Hz$ or above. Given the parameters $\mu $ and 
$\omega _{p}$ the sft depend only on the angle of inclination $\chi $. A
plot between the plasma frequency shift and the angle $\chi $; based on
expression (14) for the cases of neutron star, pulsar, and white dwarf for
the typical values of mass, radius and angular frequency is given in Fig.
(1). Clearly the gravitomagnetic reduction in th plasma frequency is maximum
in the equitorial plane of the star.

\FRAME{ftbpFU}{240.25pt}{157.6875pt}{0pt}{\Qcb{Plots for the shift $\protect%
\omega /\protect\omega _{p}$ in the plasma frequency of a compact star
atmosphere as a function of the angle of inclination $\protect\chi $ of the
plane of observation to the angular frequency vector for the case of a white
dwarf ($M=1M_{\circledast }=1.989\times 10^{30}kg$, $R=7\times 10^{6}m$, $%
\Omega =1Hz$, $\protect\omega _{p}=$\negthinspace $5.65\times 10^{2}Hz$), a
pulsar ($M=1.4M_{\circledast }$, $R=$ $3\times 10^{4}m$, $\Omega =30Hz$, $%
\protect\omega _{p}=5.65\times 10^{4}Hz$),and a neutron star ($%
M=2M_{\circledast }$, $R=$ $1\times 10^{4}m$, $\Omega =1kHz$, $\protect%
\omega _{p}=5.65\times 10^{6}Hz$).}}{\Qlb{1}}{Figure 1}{\special{language
"Scientific Word";type "GRAPHIC";display "PICT";valid_file "T";width
240.25pt;height 157.6875pt;depth 0pt;original-width
554.8125pt;original-height 329pt;cropleft "0";croptop "1";cropright
"1";cropbottom "0";tempfilename 'DKHZO200.bmp';tempfile-properties "XPR";}}

\section{Gravitomagnetic Effects on the Dielectric Constant of the Plasma}

To study the effects of plasma frequency shift on the dielectric constant of
the plasma, let $\omega _{p}^{shift}\equiv \omega _{p}-H/2$, where $H$ is
given by (9). Then in terms of plasma frequency we have the following
expression for the dielectric constant $\varepsilon $ for the plasma:

\begin{equation}
\varepsilon =1-(\frac{\omega _{p}^{shifted}}{\omega _{EM}})^{2}=1-(\frac{%
\omega _{p}}{\omega _{EM}}-\frac{\mu }{2\omega _{EM}}\sqrt{1+3\cos ^{2}\chi }%
)^{2},
\end{equation}
where $\omega _{EM}$ is the angular frequency of an electromagnetic signal
propagating through the plasma.

For an electromagnetic signal of some given angular frequency the dielectric
constant (15) depends only on the angle $\chi $ for a compact gravitational
source. Ploting this dependence of the constant $\varepsilon $ on the angle $%
\chi $ in Fig.(2) we find that the plasma medium becomes rarer as $\chi $
varies from $0$ to $\pi /2$. Therefore if $\delta \omega _{EM}$ be the
angular frequency range of the radially out-going waves, then from the usual
conditions for the propagation of an electromagnetic signal in continuous
media we have for the transmission and reflection the conditions in the
field of a rotating gravitational source:

\begin{eqnarray}
\delta \omega _{EM} &<&\omega _{p}^{shift},\quad reflection;  \nonumber \\
\delta \omega _{EM} &>&\omega _{p}^{shift},\quad transmission.
\end{eqnarray}

\ \FRAME{ftbpFU}{243.3125pt}{161.5pt}{0pt}{\Qcb{Plots for the refractive
index $\protect\varepsilon $ of a compact star atmosphere as a function of
the angle of inclination $\protect\chi $ of the plane of plasma oscillations
to the angular frequency vector for the case of a white dwarf ($%
M=1M_{\circledast }=1.989\times 10^{30}kg$, $R=7\times 10^{6}m$, $\Omega =1Hz
$, $\protect\omega _{p}=$\negthinspace $5.65\times 10^{2}Hz$), a pulsar ($%
M=1.4M_{\circledast }$, $R=$ $3\times 10^{4}m$, $\Omega =30Hz$, $\protect%
\omega _{p}=5.65\times 10^{4}Hz$),and a neutron star ($M=2M_{\circledast }$, 
$R=$ $1\times 10^{4}m$, $\Omega =1kHz$, $\protect\omega _{p}=5.65\times
10^{6}Hz$) for the propagation of an EM signal of angular frequency $10^{3}Hz
$, $10^{5}Hz$, and $10^{7}Hz$ respectively.}}{\Qlb{2}}{Figure 2}{\special%
{language "Scientific Word";type "GRAPHIC";display "USEDEF";valid_file
"T";width 243.3125pt;height 161.5pt;depth 0pt;original-width
554.8125pt;original-height 329pt;cropleft "0";croptop "1";cropright
"1";cropbottom "0";tempfilename 'DKHZRP01.bmp';tempfile-properties "XPR";}}

The shift in plasma frequency in the equatorial plane ($\chi =\pi /2$) is $%
\omega _{p}-\mu /2$ whereas in the plane orthogonal to it ($\chi =0$) the
shift is $\omega _{p}-\mu $. Denoting the frequency shift in the equatorial
plane by $\tilde{\omega}_{p\parallel }$ and by $\tilde{\omega}_{p\perp }$
for the orthogonal plane, we find that the maximum difference in the
frequencies transmitted through the plasma:

\begin{equation}
\mid \omega _{p\perp }^{shift}-\omega _{p\parallel }^{shift}\mid _{\max }=%
\frac{\mu }{2}.
\end{equation}

It is worth emphasizing here that since plasma oscillations and the
gravitomagnetic force both are along the radial direction therefore even if
the plasma co-rotates with the star (for instance close to the ion crust)
the gravitomagnetic force will effect the oscillations of the plasma and
hence the dielectric constant.

\section{Conclusions}

In this paper we have considered the effects of gravitomagnetic force on
electron oscillations in a homogeneous plasma with a fixed ion background.
It was found that there is a shift in the characteristic frequency and hence
in the dielectric constant of the plasma surrounding the star. The shift
depends on the intrinsic parameters of the star, i. e., its mass and radius,
and also on the component of the angular frequency vector of the star to the
plane of oscillation of the electron. It was estimated that an
electromagnetic wave of given frequency the shift in the dielectric constant
results in changing the allowed range of frequencies transmitted through the
plasma, by a maximum amount $\mu /2$ (as defined for expression (13) above).

For a plasma enveloping a white dwarf the predicted difference in the
frequency of electromagnetic radiation emitted is very small (the magnitude
of the parameter $\mu $ is about $0.1687\times 10^{-3}Hz$). However for a
typical pulsar ($\mu \simeq 1.6540Hz$) and especially neutron stars ($\mu
\simeq $ $236.2932Hz$) the shift, though still small, may possibly be
observed (for example in radiation spectra of various compact sources),
especially when the effects of various contingencies (such as dispersion due
to interstellar gases, effects due to the atmosphere of the Earth, etc.) are
isolated.

\end{document}